\documentclass[a4paper,fleqn,usenatbib]{mnras}

\usepackage{mathptmx}
\usepackage[T1]{fontenc}
\usepackage{ae,aecompl}

\usepackage{graphicx}	% Including figure files
\usepackage{amsmath}	% Advanced maths commands
\usepackage{amssymb}	% Extra maths symbols

\usepackage{natbib, color, graphicx}

\title[Dust properties in galactic winds from NGC~1569]{Enhanced dust
emissivity power-law index along the western H$\alpha$ filament of NGC~1569}
\author[T. Suzuki et al.]{T. Suzuki$^{1}$\thanks{E-mail: t.suzuki@u.phys.nagoya-u.ac.jp},
H. Kaneda$^{1}$, T. Onaka$^{2}$, M. Yamagishi$^{3}$, D. Ishihara$^{1}$,
T. Kokusho$^{1}$ \newauthor and T. Tsuchikawa$^{1}$ \\
$^{1}$Graduate School of Science, Nagoya University, Furo-cho
Chikusa-ku, Nagoya, 464--8602, Japan\\ 
$^{2}$Department of Astronomy, Graduate School of Science, The
University of Tokyo, 7-3-1 Hongo Bunkyo-ku, Tokyo, 113--0033, Japan \\
$^{3}$Institute of Space and Astronautical Science, Japan
              Aerospace Exploration Agency, Sagamihara,  Kanagawa
	      252--5210, Japan}

\begin{document}

%\date{Accepted 1988 December 15. Received 1988 December 14; in original form 1988 October 11}

%\pagerange{\pageref{firstpage}--\pageref{lastpage}} \pubyear{2002}

\maketitle

\label{firstpage}

\begin{abstract}
We used a data set from \textit{AKARI} and \textit{Herschel} images at
 wavelengths from 7~$\mu$m to 500~$\mu$m to catch the evidence of dust
 processing in galactic winds in NGC~1569. Images show a diffuse
 infrared (IR) emission extending from the galactic disk into the halo
 region. The most prominent filamentary structure seen in the diffuse IR
 emission is spatially in good agreement with the western H$\alpha$
 filament (western arm). The spatial distribution of the
 $F_\mathrm{350}/F_\mathrm{500}$ map shows high values in regions
 around the super-star clusters (SSCs) and towards the western arm, which
 are not found in the $F_\mathrm{250}/F_\mathrm{350}$ map. The color-color
 diagram of $F_\mathrm{250}/F_\mathrm{350}$--$F_\mathrm{350}/F_\mathrm{500}$
 indicates high values of the emissivity power-law index 
 ($\beta_\mathrm{c}$) of the cold dust component in those regions. From a
 spectral decomposition analysis on a pixel-by-pixel basis, a
 $\beta_\mathrm{c}$ map shows values ranging from 
 $\sim1$ to $\sim2$ over the whole galaxy. In particular, high
 $\beta_\mathrm{c}$ values of $\sim2$ are only observed in the regions
 indicated by the color-color diagram. Since the average 
 cold dust temperature in NGC~1569 is $\sim30$~K, $\beta_\mathrm{c}<2.0$
 in the far-IR and sub-mm region theoretically suggests emission from
 amorphous grains, while $\beta_\mathrm{c}=2.0$ suggests that from crystal grains. Given
 that the enhanced $\beta_\mathrm{c}$ regions are spatially confined by the HI
 ridge that is considered to be a birthplace of the SSCs, 
 the spatial coincidences may indicate that dust grains 
 around the SSCs are grains of relatively high
 crystallinity injected by massive stars originating from starburst
 activities and that those grains are blown away along the HI ridge and
 thus the western arm.     
\end{abstract}


\section{Introduction}
Understanding of galaxy evolution remains a key subject in modern 
astrophysics. Circulation of gas and dust on galactic scales
has a large influence on the evolution of both the interstellar medium
(ISM) and the intergalactic medium (IGM), and thus serves a vital role in
galaxy evolution.

\begin{figure*}
\centering
 \includegraphics[width=11cm]{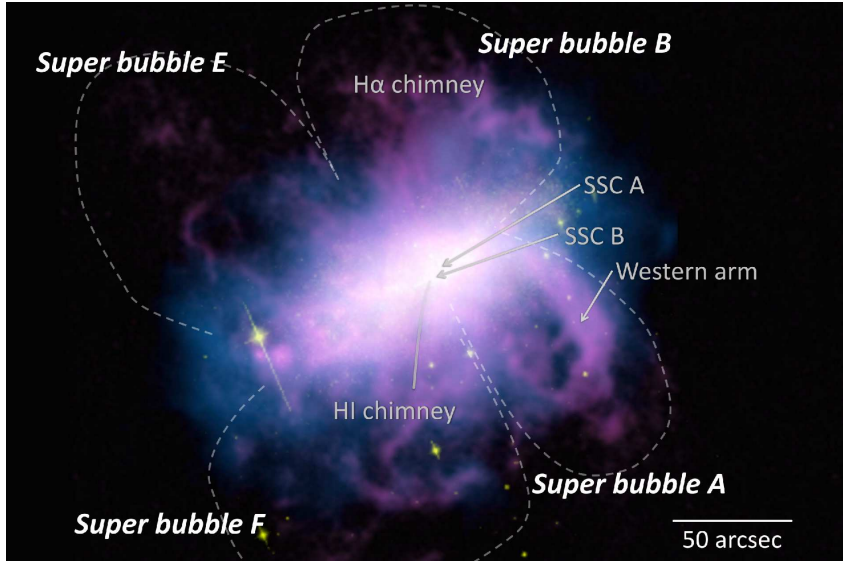}
 \caption{Composite image of NGC~1569: 0.66 $\mu$m (H{$\alpha$}) (magenta),
 0.837 $\mu$m (green) and 70 $\mu$m (cyan). \citet{Martin1998} defines
 the four-superbubble (A, B, E and F) regions denoted by the
 dashed line. H$\alpha$ filaments inside the superbubbles were
 identified by \citet{Martin1998} and \citet{Westmoquette2008}. The
 other features were identified by \citet{Hodge1974}, \citet{Arp1985},
 \citet{Israel1990}, \citet{Mule2005}, and  \citet{Hunter1993}.}
\label{fig1}
\end{figure*}

One of major drivers to eject the ISM into the IGM is
galactic winds driven by mechanical energy and momentum from
supernovae and stellar winds~\citep[e.g.,][]{Aguirre2001, Zu2011}.   
Galactic winds are observationally found at all cosmic epochs and are
very prominent and ubiquitous especially in the redshift range of
z$\sim$2--3~\citep{Erb2012}, which is the peak star formation
period. The loss of gas from galaxies can quench their star-forming
activities and consequently can change the properties of the galaxies
~\citep[e.g.,][]{Cazzoli2014}. On the other hand, the presence of 
metal-enriched gas and dust grains in the IGM directly affects the
thermal balance of gas. In particular, \cite{Montier2004} show that
infrared (IR) emission from dust grains can be considered as a dominant
cooling agent of gas with temperatures of $10^6$--$10^8$~K, which are
the typical range in the IGM; the cooling efficiency by dust grains
strongly depends on a dust-to-gas mass ratio (DGR) and a dust size
distribution. The efficient cooling by dust grains induces an increase
of star and dust formations, and may impact on hierarchical clustering
of the large-scale structures. Despite the importance of the role of dust
grains in galaxy evolutions, comprehensive understanding of the fate of
dust due to galactic winds is not observationally established yet,
although recent optical, IR and sub-millimeter~(mm) observations of
nearby starburst galaxies show extended emissions from dust grains and
polycyclic aromatic hydrocarbons~(PAHs) and revealed kinematics of PAHs
in galactic winds \citep[e.g.,][]{Kaneda2010, Roussel2010, Yoshida2011}.

The dwarf-irregular galaxy NGC~1569 is an ideal laboratory to
investigate dust processing in galactic winds and to insight into it in
the early universe. Dwarf galaxies are believed to be the `building
blocks' of larger galaxies in the early universe. In particular,
starburst dwarf irregular galaxies are expected to have experienced
gravitational interactions or mergers. NGC~1569, a member of the IC~342
group of galaxies, may have interacted with a companion
galaxy~\citep{Johnson2013} and has undergone at least three major
starburst phases; the most recent and strongest starburst ($\sim$3.2
M$_\mathrm{\sun}$ yr$^{-1}$kpc$^{-2}$) started $\sim$40 Myr ago and 
ended $\sim$10 Myr ago~\citep{Angeretti2005}. During this phase, the
galaxy witnessed $\sim10^4$ supernova explosions, and the formation of
star clusters including the two prominent super-star clusters (SSCs) A
and B~\citep{Arp1985}. These activities have had dramatic effects on
both the ISM and the halo of NGC~1569: (i) a cavity shown in the H\,{\sc i},
H$\alpha$ and far-IR distributions centered on SSC A, most
probably due to stellar winds from SSC A~\citep{Israel1990, Hunter2000,
Lianou2014}, (ii) chimneys in southward and northward directions from
SSC A, which are considered to be major pathways to eject the ISM into
the halo~\citep{Hunter1993, Mule2005, Westmoquette2008}, (iii) a
bipolar metal-enriched outflow comprising kpc-scale expanding
superbubbles on the southern and northern sides of the galactic
disc~\citep{Hunter1993, Heckman1995, Martin1998, Martin2002,
Westmoquette2008}.

\begin{table*}
 \centering
 \begin{minipage}{140mm}
  \caption{Summary of the IR and sub-mm imaging data set taken from the {\it AKARI}
  and {\it Herschel} data archives.}
  \begin{tabular}{@{}lrrrr@{}}
   \hline
  Telescope/Instrument &  $\lambda$ ($\umu$m) & FWHM (\arcsec) &
   Observation IDs & 1-$\sigma$ background noise (MJy sr$^{-1}$)\\
 \hline
  {\it AKARI}/IRC &   7 & 5.1   & 1400423 & $3.4\times10^{-2}$\\
  {\it AKARI}/IRC &  11 & 4.8 & 1400423 & $4.5\times10^{-2}$\\
  {\it AKARI}/IRC &  15 & 5.7 & 1400424 & $4.2\times10^{-2}$\\
  {\it AKARI}/IRC &  24 & 6.8 & 1400424 & $1.2\times10^{-1}$\\
  {\it Herschel}/PACS & 70 & 5.6 & 1342243816--1342243821 & 1.0\\
  {\it Herschel}/PACS & 160 & 11.4 & 1342243816--1342243821 & 1.6\\
  {\it Herschel}/SPIRE & 250 & 18.4 & 1342193013 & 1.3\\ 
  {\it Herschel}/SPIRE & 350 & 25.2 &  1342193013 & 1.5\\ 
  {\it Herschel}/SPIRE & 500 & 36.7 &  1342193013 & 0.4\\ 
\hline
\label{table1}
\end{tabular}
\end{minipage}
\end{table*}

As shown in Fig.~\ref{fig1}, H$\alpha$ observations show that the halo
contains four expanding superbubbles (A, B, E and F;
\citealt{Martin1998}), which include a number of cellular 
filaments excited by shocks; their dynamical ages are roughly estimated to be 10--25 Myr
from their expanding velocities (80--100~km~s$^{-1}$) and
diameters ($\sim$1 kpc)~\citep{Westmoquette2008}. The ages are
consistent with the starburst histories of SSC A and
B~\citep{Angeretti2005}. It indicates that each superbubble is associated
with a star-forming event. \cite{Martin2002} found the two-component
diffuse X-ray emissions in the halo region; the softer and brighter
X-ray component (0.3--0.7 keV) is found adjacent to the
H$\alpha$ filaments, while the harder X-ray component (0.7--1.1
keV) is likely spatially correlated with the centers of the
superbubbles. The softer and brighter X-ray component is associated with
shocks caused by galactic wind-halo interactions.  

At a position in the most prominent H$\alpha$ filament (western
arm), \cite{Onaka2010} found the presence of the unidentified infrared
(UIR) bands at 3.3, 6.2, 7.7 and 11.3~$\mu$m by \textit{AKARI}/IRC
spectroscopic observations. Under the shock environment, the destruction
timescale ($\sim$$10^3$ yr) of the UIR band carriers, which have been
possibly attributed to PAHs, is much shorter than the dynamical timescales of the
galactic wind in superbubble A (cf.~\citealt{micelotta2010}). The result may indicate that the band
carriers are produced by fragmentation of larger grains in shocks and
that dust processing took place over a wide area of the halo
region. Despite the possibility, extensive investigations on dust
processing in galactic winds from NGC~1569 have never been reported
because of a lack of spatially resolved far-IR images before the
\textit{Herschel} era.

In this paper, using a data set from \textit{AKARI} \citep{murakami2007}
and \textit{Herschel} \citep{Pilbratt2010} imaging observations at
wavelengths from 7~$\mu$m to 500~$\mu$m, we report spatial distributions 
of dust grains entrained by galactic winds from NGC~1569 based on
pixel-by-pixel spectral energy distribution (SED) fitting to catch the
evidence of dust processing in galactic winds.

\section{Observations and data analysis}
NGC~1569 was observed as part of the \textit{AKARI} mission program
`ISM in our Galaxy and Nearby Galaxies'~\citep[ISMGN --
P.I. Kaneda~H.;][]{kaneda2009_cospar} and as part of the \textit{Herschel}
programs `the Dwarf Galaxy Survey'~\citep[DGS --
P.I. Madden~S.;][]{Madden2013} and 'Exploring the Dust Content of
Galactic Winds with Herschel: The Dwarf Galaxy Population'~(PI:
Veilleux~S.). We used \textit{AKARI}/IRC~\citep{onaka2007} and 
\textit{Herschel}/PACS-SPIRE imaging data~\citep{Poglitsch2010,
Griffin2010}, which were taken from data archives: the \textit{AKARI}
pointing data archive through the data archive and transmission system
(DARTS) and the \textit{Herschel} Science Archive (HSA). Table~\ref{table1}
lists the summary of the data used in this study.

The IRC data obtained with the IRC02 observation mode provide mid-IR
images with four bands (7, 11, 15 and 24 $\mu$m), which are finely
allocated for probing emission from PAHs and very small grains. Each
field-of-view has a size of about $10\arcmin\times10\arcmin$. The
details of the observations are described in \cite{Onaka2010}; the FWHM of
the point spread function (PSF) ranges from
$\sim$5\arcsec~to~$\sim$7\arcsec. The mid-IR images 
with a pixel size of 2\farcs 3 were created by using the IRC imaging
pipeline software version 20131202; overall four-band images thus
obtained are the same as those shown in Fig.~1 of \cite{Onaka2010}. However,
for the four-band mid-IR images, in particular the IRC 24 $\mu$m
image, mid-IR emission in the halo region of interest is affected by an
extended component of the PSF due to diffraction and scattering when
bright sources such as SSCs are observed. To correct for the extended
PSF effects, the image reconstruction method proposed by
\cite{Arimatsu2011} was applied for the four-band images; the method
consists of the deconvolution with the intrinsic PSF of an input image
and of the convolution with a Gaussian pattern with the same FWHM of the
PSF. To apply the convolution, the pixel size of the four-band images
was reduced by a factor of 2 (1\farcs 15). Furthermore, the point
sources located in the halo region were removed with the 
following procedures: first, a point source was fitted with a model of
2-D gaussian plus constant components so that the gaussian center is
located at the peak brightness of the point source. The radius of each
fitting area is set to be three times as large as the FWHM. Second, the
gaussian component was subtracted from the original image. Finally, for
subtracted pixels, values below the mean surface brightness given by the
constant component were replaced by the mean surface brightness. A
series of procedures was applied to each point source. The systematic
flux uncertainty, which is dominantly caused by the PSF reconstruction,
is estimated to be 7.4\% (IRC 7~$\mu$m), 7.5\% (IRC 11~$\mu$m), 18\%
(IRC 15~$\mu$m) and 16\% (IRC 24~$\mu$m).

Cross-linked observations were performed by the PACS, and
simultaneously provide 70 and 160 $\mu$m-band images which cover an area
of $\sim$$7\arcmin \times 9\arcmin$ around
NGC~1569. Both data were taken from the HSA and are
level-3 products~(SPG version 14.2.0) providing Unimap maps.
The FWHM of the PSF is 5\farcs
6~for PACS~70~$\mu$m and 11\farcs 4~for PACS~160~$\mu$m. 
The systematic flux calibration uncertainty is 5\% for
both bands (PACS Observer's Manual version~2.5.1, 2013).

\begin{figure*}
\vspace{0mm}
\hspace{-3.5mm}
\centering
 \includegraphics[width=18.cm]{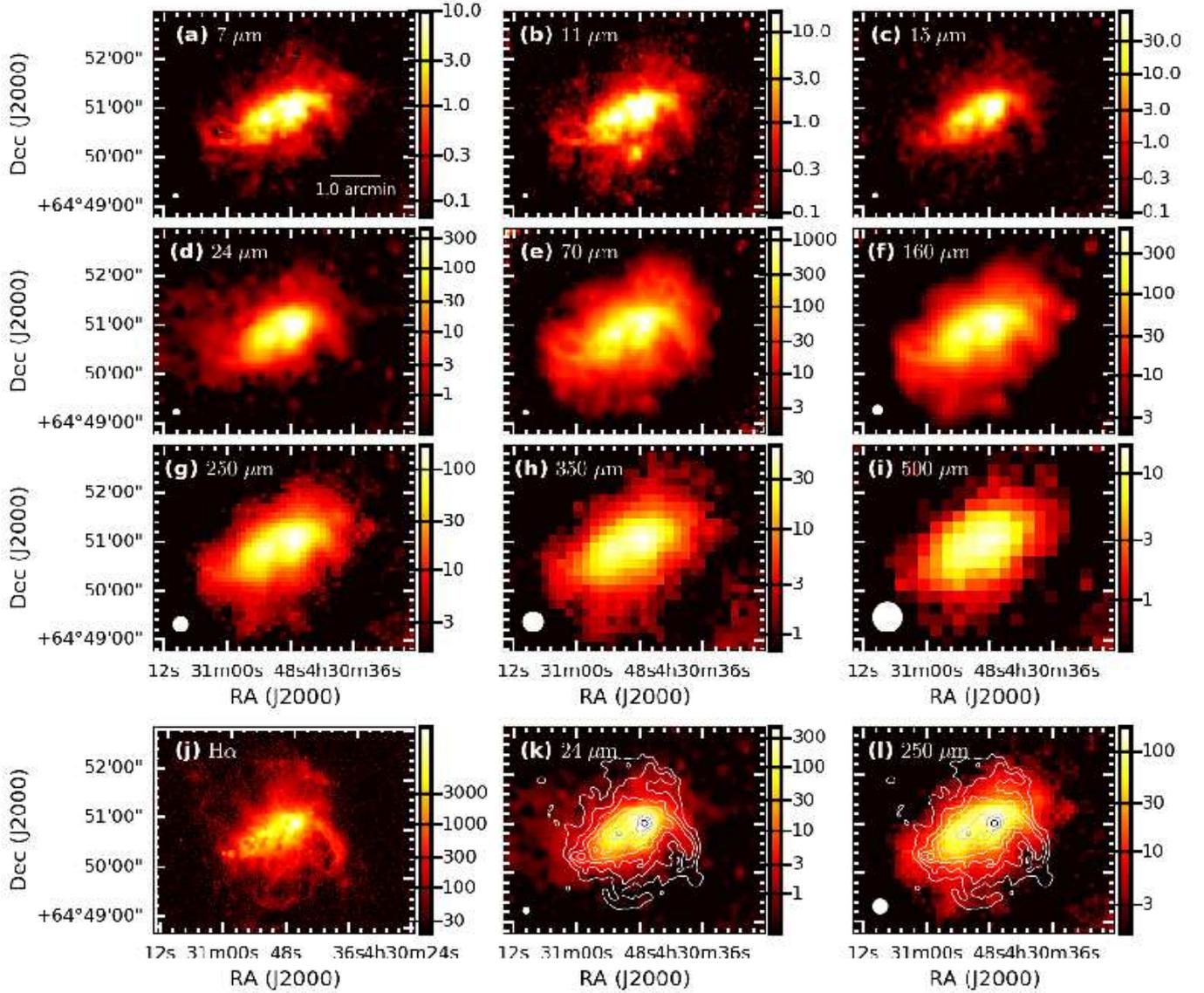}
 \caption{Background-subtracted nine-band images of NGC~1569 with the
 \textit{AKARI}/IRC and \textit{Herschel}/PACS-SPIRE in the (a)
 IRC~7~$\mu$m, (b) IRC~11~$\mu$m, (c) IRC~15~$\mu$m, (d) IRC~24~$\mu$m,
 (e) PACS~70~$\mu$m, (f) PACS~160~$\mu$m, (g) SPIRE~250~$\mu$m, (h)
 SPIRE~350~$\mu$m and (i) SPIRE~500~$\mu$m bands. In each image, the
 color bar is given in units of MJy sr$^{-1}$. The PSF size in FWHM is
 shown in the lower left-hand corner. For comparison with the spatial
distribution of the H$\alpha$ emission shown in the
panel~(j)~\citep{Hunter2004}, the same panels of (d) and (g) are shown
together with the H$\alpha$ contours in panels (k) and (l).}    
\label{fig2}
\end{figure*}

SPIRE large cross-scan observations provide 250, 350 and 500 $\mu$m-band
images at the same time and cover an area of $\sim$ $20\arcmin \times
20\arcmin$ around NGC~1569 in each band. The details of the observations are
described in \cite{Remy2013}. The SPIRE data were taken
from the HSA and are level-2 products (SPG version 14.1.0) providing
\textit{Naive} maps with the pixel sizes of 6\arcsec for
SPIRE~250~$\mu$m, 10\arcsec for SPIRE~350~$\mu$m and 14\arcsec for
SPIRE~500~$\mu$m. The beam size has the FWHMs of 18\farcs 4, 25\farcs 2
and 36\farcs 7 at SPIRE~250~$\mu$m, SPIRE~350~$\mu$m and
SPIRE~500~$\mu$m, respectively. The systematic flux calibration
uncertainty of the three SPIRE bands is applied to be $\sim$5\%~(SPIRE
Observer's Manual version~2.5, 2014).

A background level was estimated by averaging the values from multiple
apertures placed around 2\arcmin--3\arcmin away from the galaxy center without
overlapping with extended emission and was subtracted from each image. 
For multi-band analysis, the spatial resolutions of
background-subtracted IR images were reduced to match the PSF of the
SPIRE~500~$\mu$m data by convolving PACS and SPIRE-band
images with kernels provided by~\cite{Aniano2011}. Because such kernels
for the AKARI bands are currently not available in the
library\footnote{http://www.astro.princeton.edu/$\sim$ganiano/Kernels.html}, 
the AKARI images were convolved by a Gaussian kernel to approximate the
SPIRE~500~$\mu$m PSF by the \textit{moderate} Gaussian FWHM of
41\arcsec~\citep{Aniano2011}. Then, the convolved images were
regridded with a pixel size of 36\farcs7 matched with the PSF 
of the SPIRE~500~$\mu$m data.

%\begin{figure*}
%\centering
% \includegraphics[width=14cm, clip]{./fig3.ps}
% \caption{ISRF map from a GALEX FUV image. The color bar is in units of the Habing
%field ($G_0$=$2.3\times10^{-3}$ erg s$^{-1}$ cm$^{-2}$ at
% $\lambda=1530$~\AA). The contours superimposed on the image show the
% H$\alpha$ emission, the same as those in Fig.~\ref{fig2}. }
%\label{fig3}
%\end{figure*}

\begin{figure*}
\vspace{-0mm}
\hspace{-3mm}
\centering
 \includegraphics[width=18cm, clip]{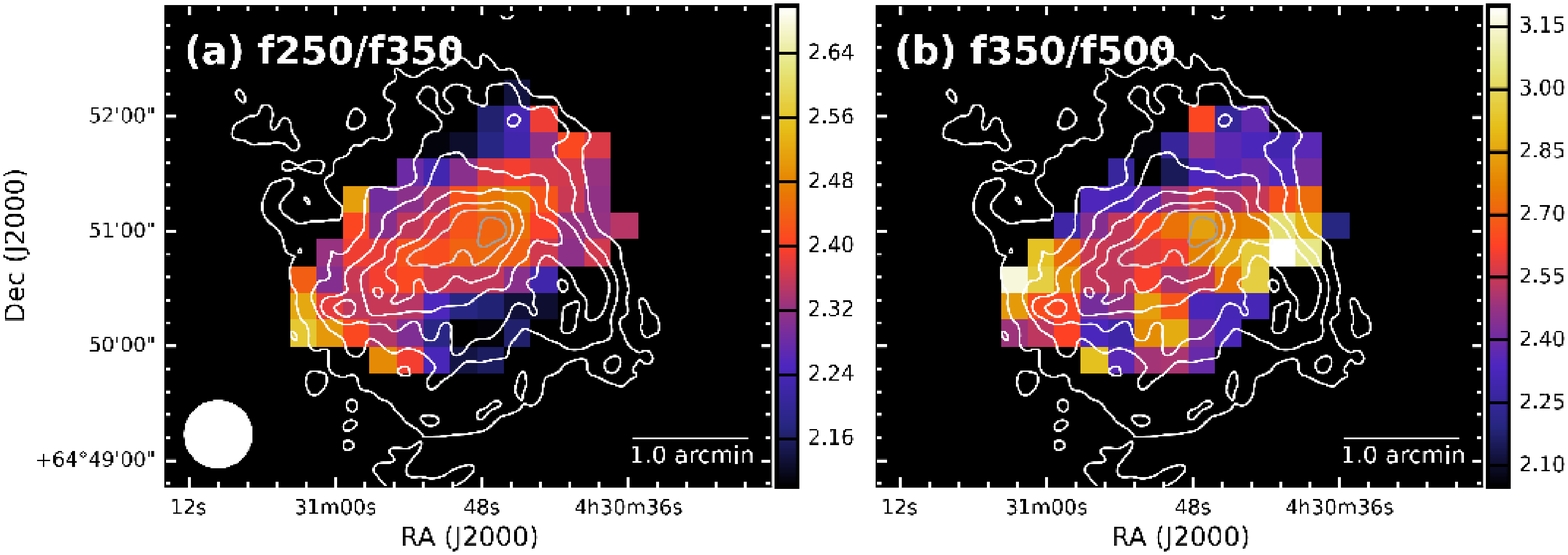}
 \includegraphics[width=18cm, clip]{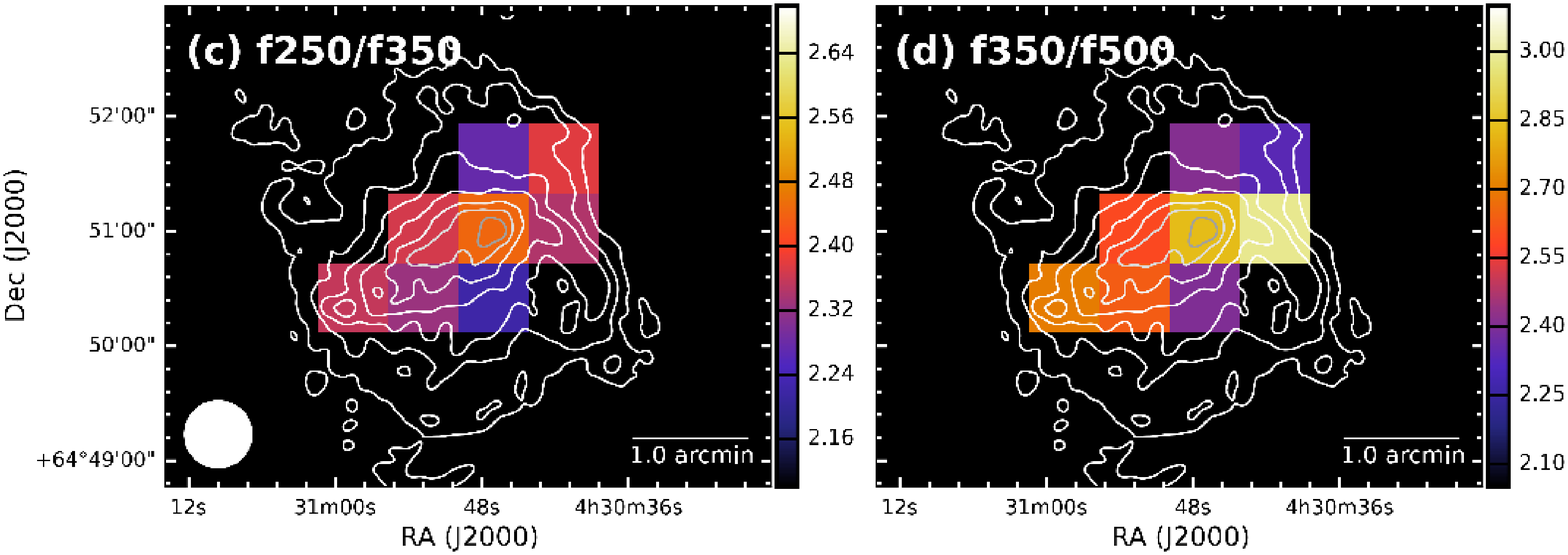}
 \caption{Color maps of (a)
 \textit{F}$_{250}$/\textit{F}$_{350}$, and (b)
 \textit{F}$_{350}$/\textit{F}$_{500}$ with the original pixel size of
 the SPIRE~500~$\mu$m data (14\arcsec). Panels (c) and (d) are the same
 as (a) and (b), but with a pixel size of 36\farcs 7,
 respectively. The PSF shown in the panels (a) and (c) is matched to the
 SPIRE 500~$\mu$m-band PSF. The detection threshold is set to be the
 5$\sigma$ level for each band image. The contours superimposed on the
 images are the same as those in Fig.~\ref{fig2}.}    
\label{fig4}
\end{figure*}

\section{Results}
\subsection{IR images}
Figure~\ref{fig2} shows the background-subtracted images of NGC~1569
obtained with \textit{AKARI} and
\textit{Herschel}. To compare them with the spatial
distribution of the H$\alpha$ emission shown in the
panel~(j)~\citep{Hunter2004}, the same panels of (d) and (g) are shown
together with the H$\alpha$ contours in panels (k) and (l). The 1-$\sigma$ background
fluctuations per pixel are 0.04--0.1~MJy sr$^{-1}$ for 7--24~$\mu$m and
0.4--1.6~MJy sr$^{-1}$ for 70--500~$\mu$m~(Table~\ref{table1}). Most of the images
clearly show diffuse IR emission extending from the galactic disk into the halo region
(hereafter referred to as IR-halo emission). Filamentary structures seen
in the H$\alpha$ emission are spatially correlated with those
seen in the IR-halo emission. In superbubbles~A and E, a pair of filaments
which outline the western and eastern sides of each superbubble is observed
in mid-IR to sub-mm images; the most prominent IR filament is spatially in
good agreement with the western arm. In
superbubble~B, the IR-halo emission extends toward north along the
H$\alpha$ chimney. With regard to superbubble~F, which is formed by the ISM
escaping from the galactic disk through the H\,{\sc i}
chimney~\citep{Westmoquette2008}, patchy IR-halo emission is
observed along H$\alpha$ filaments. Furthermore, the
IR-halo emission also shows the component extending in the major-axis
direction over the H$\alpha$ emission.  

The IRC~7~$\mu$m and 11~$\mu$m images are dominated by the PAH
emission not only from the galactic disk itself but also from the halo;
\cite{Onaka2010} detected the PAH emission from the western arm. Overall
spatial distributions in the 7~$\mu$m and 11~$\mu$m images are in good
agreement with those in the 160--500~$\mu$m images which trace the
spatial distribution of the cold dust component (20--30~K,~\citealt{Remy2013,
Lianou2014}), associated mainly with the old stellar population. The fact may
indicate that PAHs coexist with cold dust. The IRC~15~$\mu$m and
24~$\mu$m images trace emission from both hot (100--200~K) and warm
(40--60~K) dust components~\citep{Lianou2014, Galliano2003}, 
associated mainly with massive star-forming regions. Therefore, the spatial
distributions of cold and PAH components are more extended toward the
halo region than those of hot and warm dust components.

 \begin{figure*}
 \includegraphics[width=13.5cm, clip]{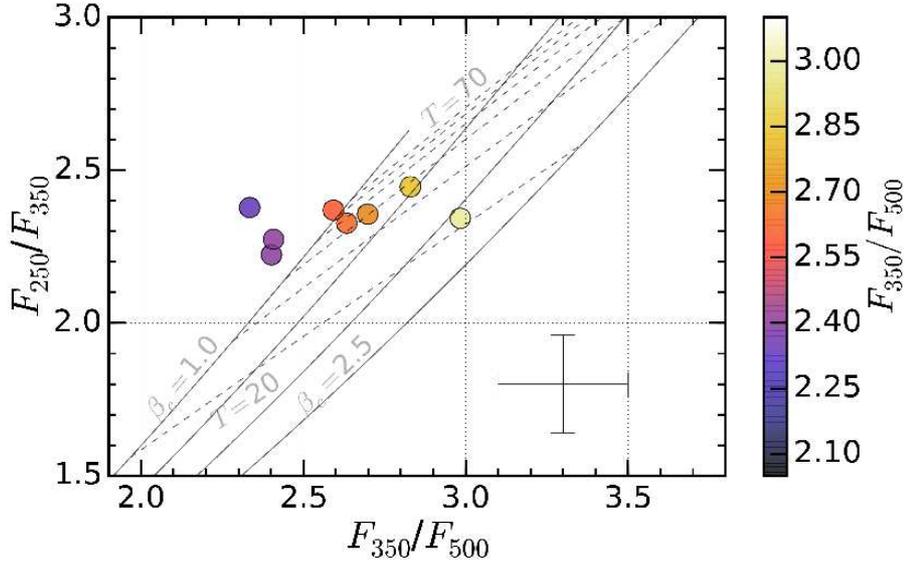}
 \vspace{-2mm}
 \caption{\textit{F}$_{250}$/\textit{F}$_{350}$--\textit{F}$_{350}$/\textit{F}$_{500}$
 diagram. The filled circle represents the observed color-color diagram at
 each pixel shown in Figs.~\ref{fig4}c and \ref{fig4}d, while the solid lines
  are the color-color diagram based on a single-component modified blackbody model defined as 
 $\nu^{\beta_\mathrm{c}} B_\nu(T_\mathrm{c})$ for
 $\beta_\mathrm{c}$=1.0--2.5 with a step of 0.5. The color of each data
  point is the same as that of each pixel in Fig.~\ref{fig4}d. 
The constant temperature points are connected by
the dashed lines for temperatures from 20 (bottom line) to 70~K
(top line) with a step of 10~K. A typical error bar is shown in
 the lower right corner.}   
\label{fig5}
\end{figure*}

\subsection{Color-color diagram:
   \textit{F}$_{250}$/\textit{F}$_{350}$--\textit{F}$_{350}$/\textit{F}$_{500}$}
To obtain a global view of the variation in the shape of the cold dust
component in the local SEDs, sub-mm colors are a simple and useful
quantitative tool to avoid potential contamination from the warm dust
component~\citep{Bendo2012, Bendo2015}. As demonstrated 
by~\cite{Boselli2012}, we created \textit{F}$_{250}$/\textit{F}$_{350}$
and \textit{F}$_{350}$/\textit{F}$_{500}$ color maps. The errors on the flux
densities ($\sigma$) were calculated from
$(\sigma^2_\mathrm{sys}+\sigma^2_\mathrm{stat} )^{1/2}$, where 
$\sigma_\mathrm{sys}$ and $\sigma_\mathrm{stat}$ are systematic and
statistical uncertainties, respectively. 

In Figs.~\ref{fig4}a and \ref{fig4}b that were regridded with the
original pixel size of the SPIRE~500~$\mu$m data (14\arcsec) for a
display purpose, values of the \textit{F}$_{250}$/\textit{F}$_{350}$
show almost constant over the whole region, while
\textit{F}$_{350}$/\textit{F}$_{500}$ values are higher in the southern
halo region, in particular the western arm, than in the northern halo
region. Those trends can clearly be confirmed in the sub-mm color-color
diagram. The filled circles in Fig.~\ref{fig5} represent the observed color-color
diagram at each pixel of the \textit{F}$_{250}$/\textit{F}$_{350}$ and
\textit{F}$_{350}$/\textit{F}$_{500}$ maps~(Figs.~\ref{fig4}c and
\ref{fig4}d). The solid lines are the 
model-predicted color-color diagram based on a single-component  
modified blackbody model with the emissivity power-law index (hereafter
referred to as the emissivity index) of 1.0 (leftmost
line), 1.5, 2.0, and 2.5 (rightmost line). The constant temperature points are
connected by the dashed lines for temperatures from
20~K (bottom line) to 70~K (top line) with a step of 10~K.
The color-color diagram suggests that the plots in and around the disk
region tend to distribute along the direction of the temperature change
with $\beta \sim1$, while those in the IR-halo region tend to
distribute relatively along the constant temperature lines 
 (dashed lines) rather than the constant $\beta$ lines; the western arm
 region shows higher $\beta$~($\sim2$), while the northern 
 halo region of the SSCs shows lower $\beta$~($\la1$). 
Since it is possible that color values are affected by
noise as a limitation of the color-color analysis, further confirmation
of the $\beta$ variation is required by performing SED fitting.

\subsection{Spectral decomposition into cold dust, warm dust and PAH components}
Spectral decomposition analysis on a pixel-by-pixel basis provides
investigations on the spatial distributions of dust properties such as dust
temperature and emissivity index. 
An individual SED constructed from the nine-band fluxes at each pixel
is reproduced by a double-component modified blackbody plus a PAH model
expressed as  
\begin{equation}
 \begin{split}
 F_{\nu,\mathrm{IR}} &= A_\mathrm{PAH}F_{\nu,\mathrm{PAH}} + A_\mathrm{c} \nu^{\beta_\mathrm{c}} B_{\nu}(T_\mathrm{c})\\
  &\quad + A_\mathrm{w} \nu^{\beta_\mathrm{w}}
 B_{\nu}(T_\mathrm{w})\left\{1+ \left(\frac{\nu_c}{\nu}\right)^\alpha e^{-\left(\frac{\nu_c}{\nu}\right)^2} \right\}, 
\label{eq1}
 \end{split}
\end{equation}
where $T_\mathrm{c}$, $T_\mathrm{w}$, $\beta_\mathrm{c}$,
$\beta_\mathrm{w}$, $A_\mathrm{c}$, $A_\mathrm{w}$, $A_\mathrm{PAH}$,
and $B_\nu(T)$ are the temperatures of cold and warm dust, the dust
emissivity indeces of cold and warm dust, the amplitudes of cold dust, 
warm dust, PAH components, and the Planck function, respectively.  
The second term in the warm dust component is the
analytic approximation of dust emission with different dust temperatures
assuming a power-law temperature distribution to take the hot dust
component into account in mid-IR
wavelengths~\citep{Casey2012}. The power-law turnover frequency
$\nu_c$ defined by~\cite{Casey2012} is a function of the mid-IR power-law
slope $\alpha$ and $T_\mathrm{w}$.
The flux density of the PAH component, $F_\mathrm{PAH}(\nu)$, is
calculated as described in \cite{suzuki2010} and is based on
the PAH parameters taken from \cite{li} and \cite{li2} by assuming
the PAH size distribution ranging from 3.55 to 300 \AA, the fractional
ionization and the temperature probability distribution for the typical
diffuse ISM with the interstellar radiation field in the solar
neighborhood.
PAHs with sizes larger than 15~\AA~contribute to $\sim20$~$\mu$m continuum
emission~\citep{li2}. Since very small grains (VSGs) which are
stochastically heated by absorbed photons contribute to $\sim20$~$\mu$m
continuum emission, the PAH component in Eq.~(\ref{eq1}) takes the VSG
emission into account.

\begin{figure*}
\centering
 \includegraphics[width=14cm, clip]{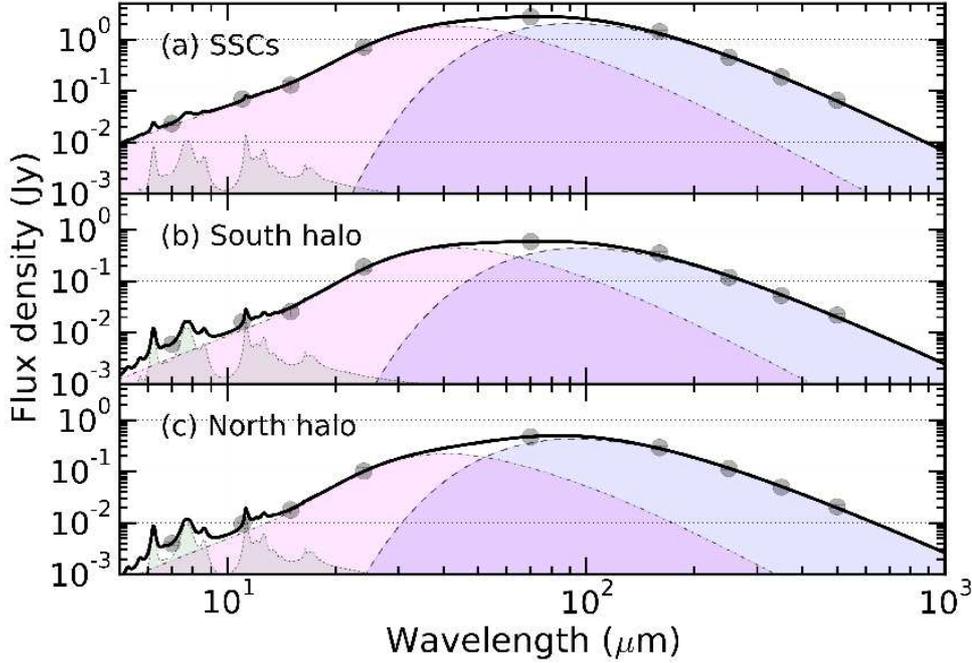}\
\vspace{-0mm}
 \caption{SEDs at pixel (36\farcs7) positions of (a) SSCs, (b)  south halo, and (c)
 north halo regions as denoted by open circles in Fig.~\ref{fig7}a. At each
 panel, the solid thick line shows the best-fit model with
 $\beta_\mathrm{w}=2.0$, which is described in Eq.~(\ref{eq1}). The cold
 dust, warm dust, and PAH components are indicated by dashed,
 dash-dotted, and dotted lines, respectively. The PSF is matched to the
 SPIRE 500~$\mu$m-band PSF. }   
\label{fig6}
\end{figure*}

\subsubsection{Pixel-by-pixel SED fitting}
Since Fig.~\ref{fig5} indicates the $\beta_\mathrm{c}$ variation over
the disk-IR halo region, $\beta_\mathrm{c}$ is set to be free, while
$\beta_\mathrm{w}$ is assumed to be constant over the disk-IR halo
region. To confirm that the assumption does not affect the result of the
$\beta_\mathrm{c}$ variation, we performed pixel-by-pixel SED fitting
for each of $\beta_\mathrm{w}$=1.0, and 2.0. 
In general, when $T_\mathrm{d}$ and $\beta$ are treated as fitting
parameters, a $\beta$--$T_\mathrm{d}$ anti-correlation is
expected~\citep[and references therein]{Tabatabaei2014}. The 
anti-correlation mainly comes from $\beta$--$T_\mathrm{d}$
degeneracy; for a set of data points, a high goodness of fit can be
obtained for different ($\beta$, $T_\mathrm{d}$) pairs. 
To avoid the artificial anti-correlation, 
one fitting cycle consists of the following two steps: the first fitting
step with fixed $T_\mathrm{c}$ and free $\beta_\mathrm{c}$, and second
fitting step with free $T_\mathrm{c}$ and fixed $\beta_\mathrm{c}$. A
series of fitting cycles for each pixel was performed and was completed
when $\chi^2$ converges with the lowest value. Thus, at each pixel, the
six-free parameters ($T_\mathrm{c}$ or $\beta_\mathrm{c}$,
$T_\mathrm{w}$, $A_\mathrm{c}$, $A_\mathrm{w}$, $A_\mathrm{PAH}$, and $\alpha$) are
derived from the nine bands. 

\begin{figure*}
\centering
 \includegraphics[width=17.5cm, clip]{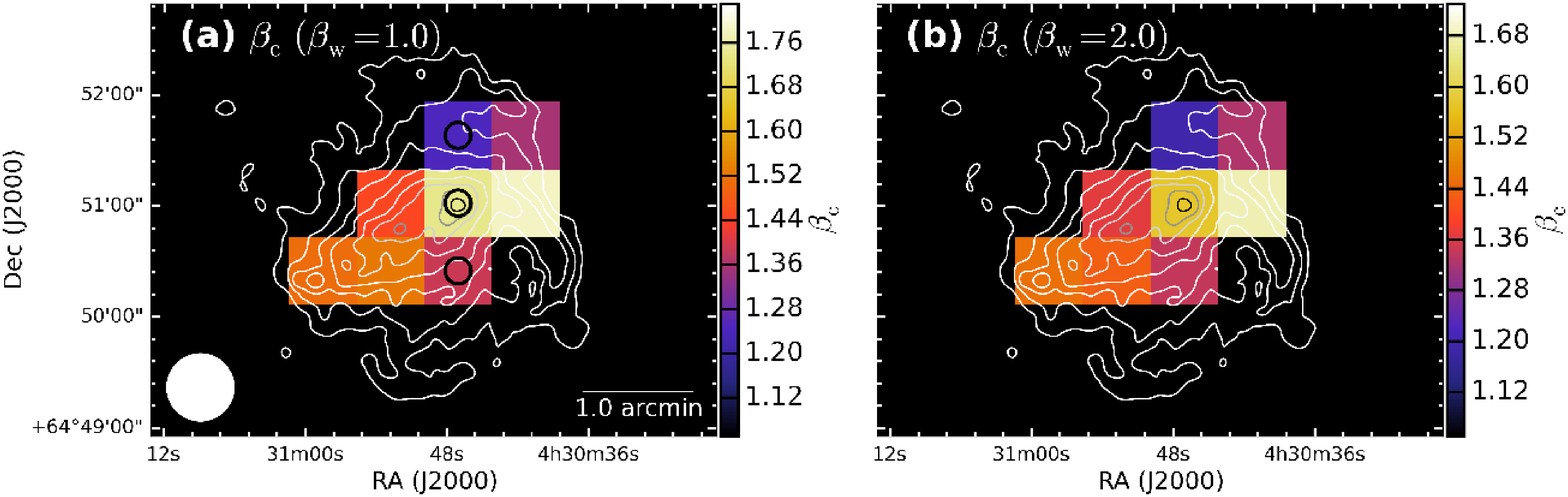}
 \includegraphics[width=17.5cm, clip]{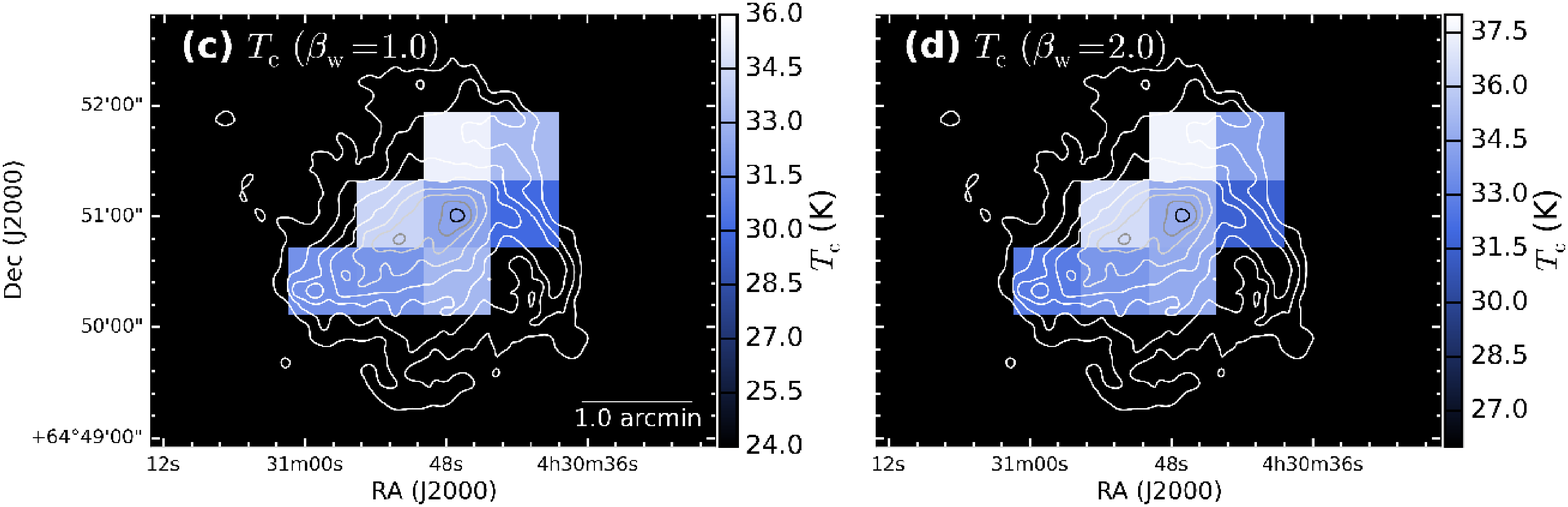}
 \includegraphics[width=17.5cm, clip]{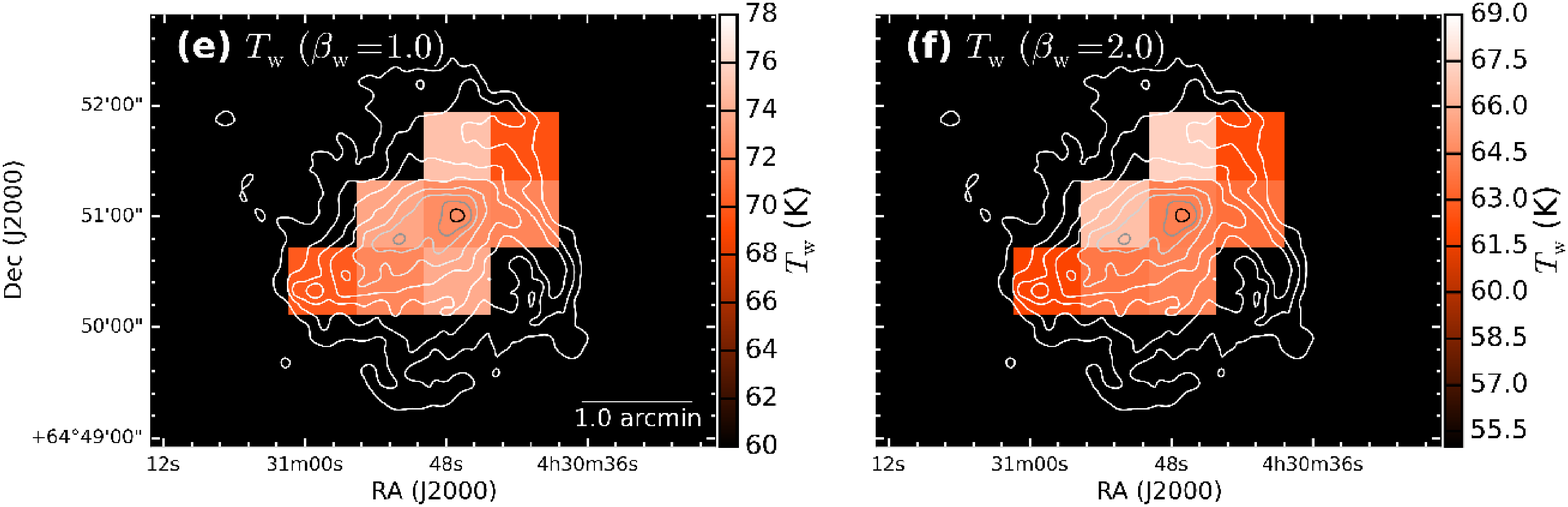}
\vspace{-3mm}
 \caption{Spatial distributions of $\beta_\mathrm{c}$ (top),
 $T_\mathrm{c}$ (middle), and $T_\mathrm{w}$ (bottom). In each column,
 left and right panels show their maps  with $\beta_\mathrm{w}=1.0$ and
 2.0, respectively. The open circles in the panel (a) 
 represent pixel positions of SEDs in SSCs, south halo, and north halo regions
 (see Fig.~\ref{fig6}). The detection
 threshold is set to be the 5$\sigma$ level for each band image.
The contours superimposed on the
 images are the same as those in Fig.~\ref{fig2}. The PSF size in SPIRE
 500~$\mu$m is shown in the lower left-hand corner on the panel (a). The
 pixel size is the same as the original one of the SPIRE~500~$\mu$m data
 (36\farcs 7).}
\label{fig7}
\end{figure*}

To minimize the probability of falling into local minima of the merit function, $\Sigma
(F_{\nu_\mathrm{model}}-F_{\nu_\mathrm{obs}})^2/\sigma^2$,
where $\sigma$ is the flux uncertainty, we generated uniform
samples of $10^6$ random sets of the six parameters. As for each
parameter, the random value ranges within 5\% from the input value which
is obtained from the previous fitting step.  
For the first fitting cycle, input values of $T_\mathrm{c}$ and
$T_\mathrm{w}$ were set to be the values obtained from SED fitting of
the whole galaxy with $\beta_\mathrm{c}=1.5$ and $\beta_\mathrm{w}=1.0$
or 2.0.   
From each random set of the six parameters, $F_{\nu,\mathrm{IR}}$ at each wavelength
was calculated. For each pixel, the initial parameter set which provides
the fluxes most closest to those observed was chosen by maximizing the
likelihood. Then, using the initial parameter set, pixel-by-pixel SED
fitting was performed to minimize $\chi^2$. Color corrections to the
IRC~15~$\mu$m, 24~$\mu$m, PACS and SPIRE-band fluxes were iteratively
performed with the assumption of a modified blackbody spectrum for each
fitting cycle. As a result, the six
parameters with $\beta_\mathrm{w}$=2.0 better reproduce the observed 
fluxes with pixel-averaged $\chi^2_\nu~(d.o.f=3)$ of
0.6. Figure~\ref{fig6} shows SEDs at pixel positions of (a) SSCs, (b) south  
halo, and (c) north halo regions, which are denoted by open circles in
Fig.~$\ref{fig7}$a, together with the best-fit model.

\begin{figure*}
\centering
 \includegraphics[width=17.5cm, clip]{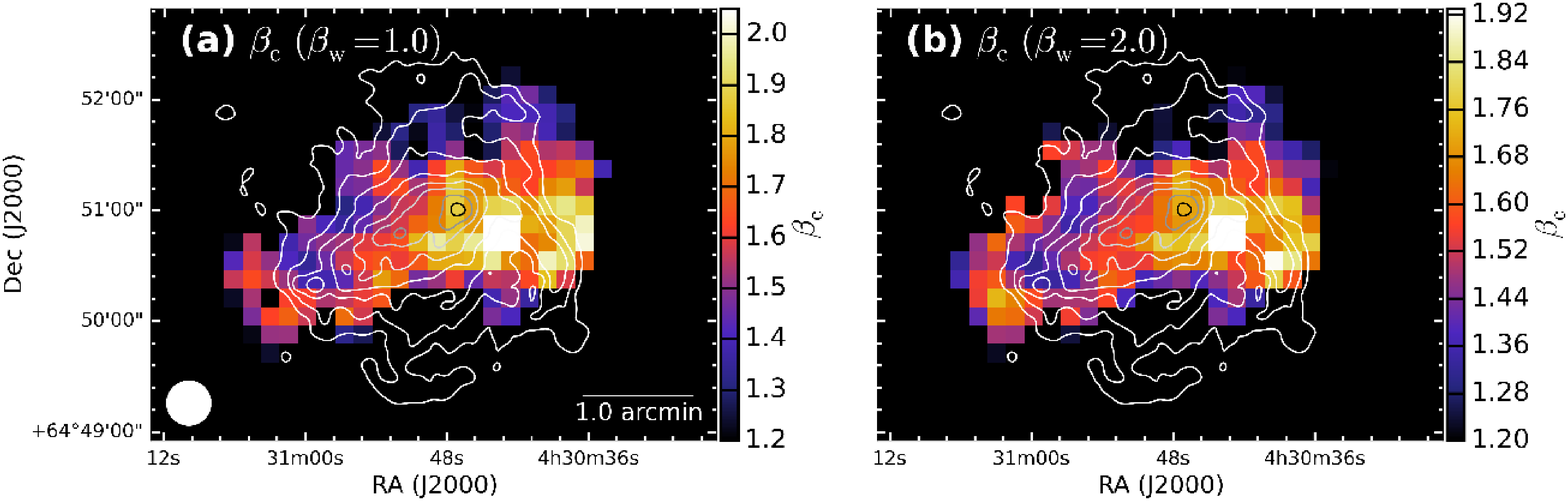}
 \includegraphics[width=17.5cm, clip]{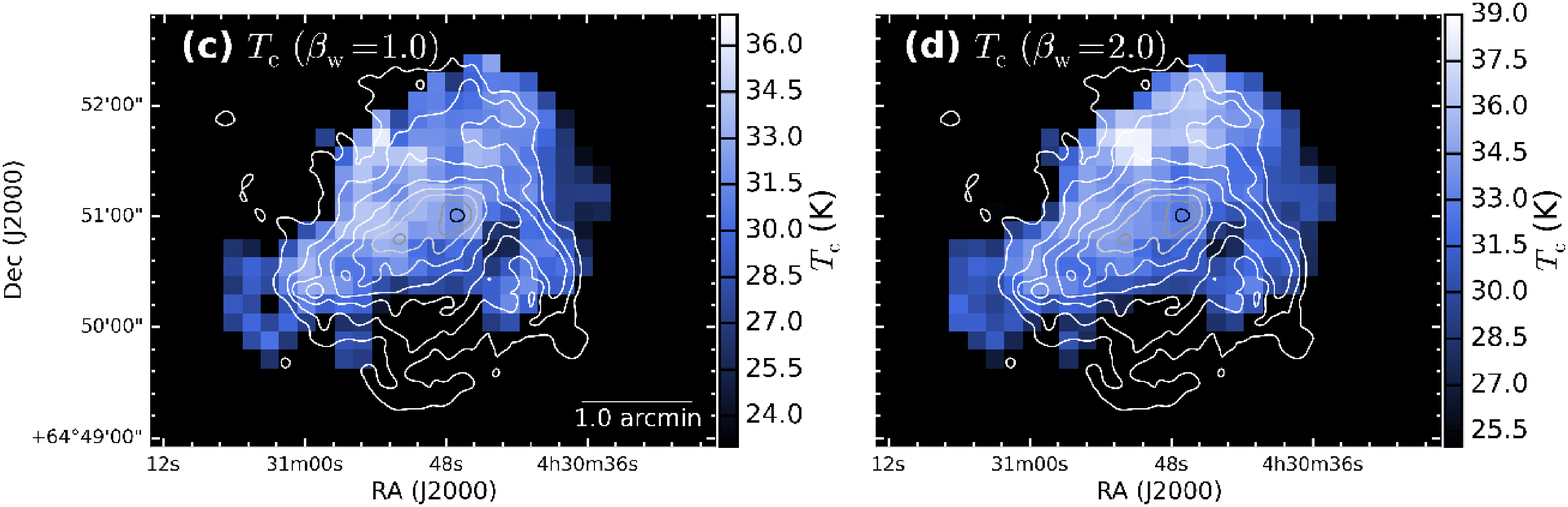}
 \includegraphics[width=17.5cm, clip]{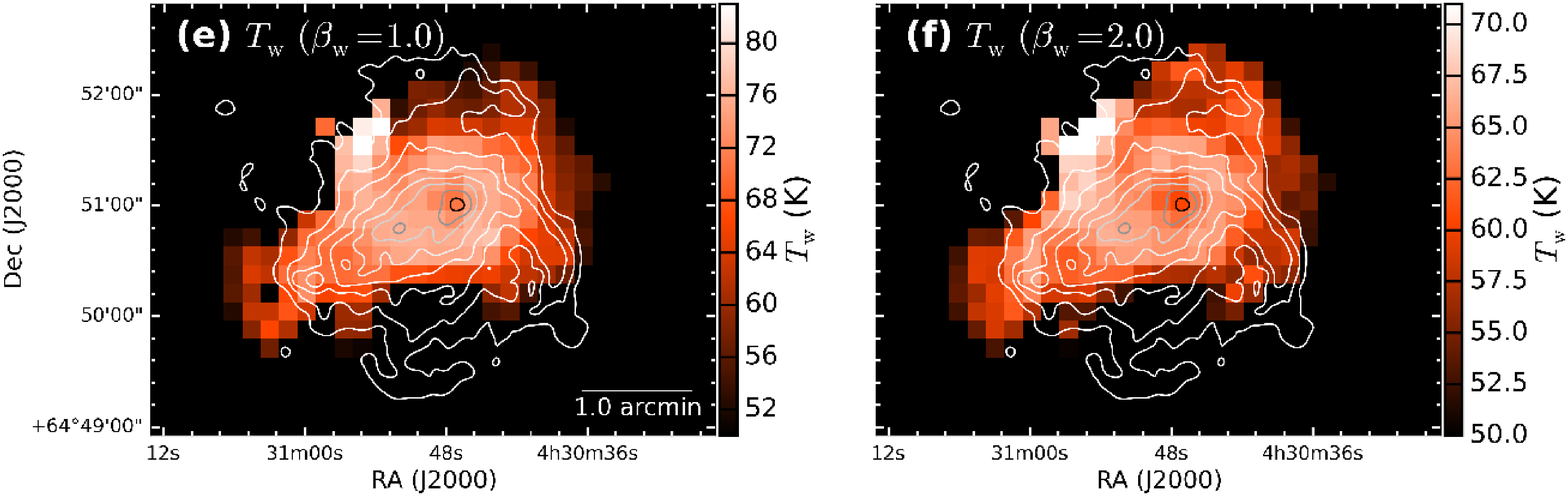}
\vspace{-3mm}
 \caption{Same as Fig.~\ref{fig7}, but the maps obtained from
 SED fitting using the eight-band images (IRC~7~$\mu$m--SPIRE
 350~$\mu$m) to show the maps with a higher 
 spatial resolution matched to the SPIRE~350~$\mu$m-band PSF (25\farcs 2). The pixel
 size is the same as the original one of the SPIRE~350~$\mu$m data (10\arcsec).} 
 \label{fig77}
\end{figure*}

\subsubsection{Spatial distributions of $\beta_\mathrm{c}$, $T_\mathrm{c}$, and $T_\mathrm{w}$}
Figure~\ref{fig7} shows the spatial distributions of $\beta_\mathrm{c}$
(top), $T_\mathrm{c}$ (middle), and $T_\mathrm{w}$ (bottom). The panels
in the left and right columns show their maps obtained with
$\beta_\mathrm{w}$ of 1.0, and 2.0, respectively. Moreover, Fig.~\ref{fig77}
shows the maps with a higher spatial resolution obtained from SED fitting
using the eight-band images~(IRC~7~$\mu$m--SPIRE~350~$\mu$m), whose PSFs
were matched to the SPIRE~350~$\mu$m-band PSF. Uncertainties
in $\beta_\mathrm{c}$, $T_\mathrm{c}$, and $T_\mathrm{w}$ 
are 4\%, 5\%, and 6\%, respectively. From Figs.~\ref{fig7}
and \ref{fig77}, it is confirmed that overall spatial distributions of
$\beta_\mathrm{c}$, $T_\mathrm{c}$, and $T_\mathrm{w}$ are independent of the
$\beta_\mathrm{w}$ values. 

\begin{figure*}
\centering
 \includegraphics[width=12cm, clip]{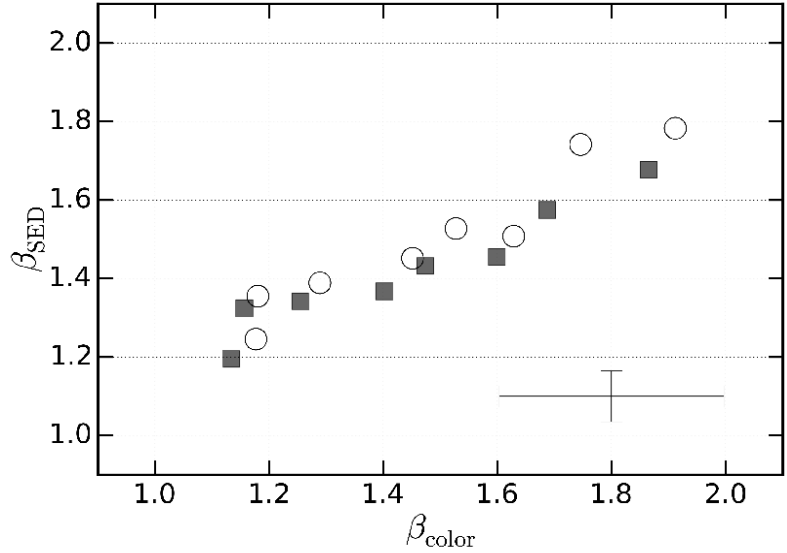}
\vspace{-3mm}
 \caption{Relation between $\beta_\mathrm{c}$ from the
 $F_\mathrm{350}/F_\mathrm{500}$ color ($\beta_\mathrm{color}$) and $\beta_\mathrm{c}$ from the
 SED fitting ($\beta_\mathrm{SED}$) with $\beta_\mathrm{w}=1.0$ (circle) and 2.0 (square)
 cases. The $\beta_\mathrm{c}$ values are obtained based on the SPIRE
 500~$\mu$m-band PSF with a pixel size of 36\farcs 7. A typical error
 bar is shown in the lower right corner.}  
\label{fig8}
\end{figure*}

As suggested by the color-color diagram, Fig.~\ref{fig7}
shows that high $\beta_\mathrm{c}$ values are observed around the western arm and 
SSCs regions. In fact, the pixel-by-pixel comparison of
$\beta_\mathrm{c}$ in Fig.~\ref{fig8} shows that the $\beta_\mathrm{c}$
values obtained from SED fitting ($\beta_\mathrm{SED}$) are in agreement
with those from the $F_\mathrm{350}/F_\mathrm{500}$ color
($\beta_\mathrm{color}$), which are obtained with a single-component
modified blackbody model. The result suggests that $\beta_\mathrm{SED}$
is not influenced by the warm dust component.
The $\beta_\mathrm{c}$ map with a higher spatial resolution (25\farcs 2,
Fig.~\ref{fig77}) reveals that $\beta_\mathrm{c}$ shows almost constant value of $\sim$1.5 in the disk
region, but high $\beta_\mathrm{c}$ values as high as $\sim2$ are found
around the SSCs and in regions towards the western arm. As for the
IR-halo region, enhanced $\beta_\mathrm{c}$ values ($\sim2$) are clearly
observed along the western arm in superbubble~A, while such high
$\beta_\mathrm{c}$ values are not observed in other three superbubbles;
the lowest $\beta_\mathrm{c}$ values are observed in superbubble~B. 

The $T_\mathrm{w}$ map in Fig.~\ref{fig77} is in agreement with the
H$\alpha$ map in the disk region. 
High-$T_\mathrm{w}$ values are observed near current star-forming regions
such as the SSCs, while the $T_\mathrm{c}$ 
map shows a more uniform distribution than the $T_\mathrm{w}$ map;
high-$T_\mathrm{c}$ values are found around SSCs. Those results
are consistent with the spatial distribution of $T_\mathrm{c}$ shown in
\cite{Lianou2014}. In the IR-halo 
region, both $T_\mathrm{c}$ and $T_\mathrm{w}$ are found to be systematically
higher on the northern side than on the southern side. Those
high-temperature dust grains tend to be spatially matched adjacent to the
H$\alpha$ filaments of superbubbles B and E.

\begin{figure*}
\centering
 \includegraphics[width=12cm, clip]{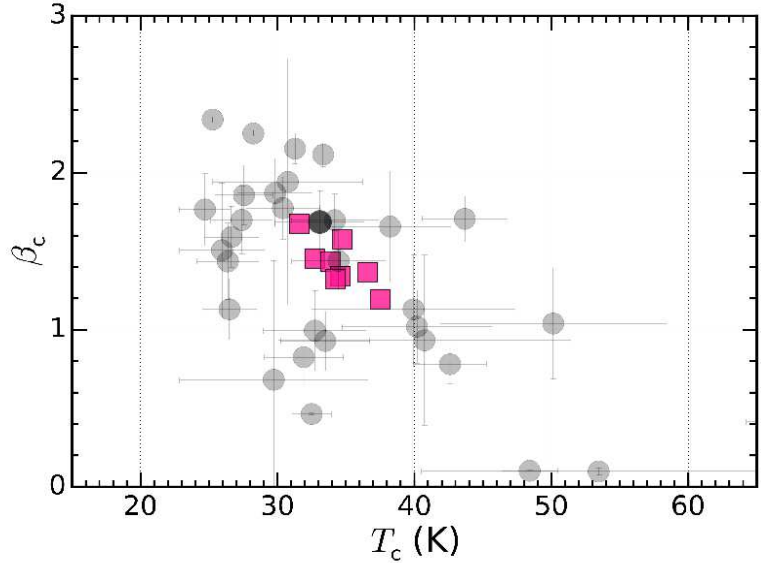}
\vspace{-3mm}
 \caption{Pixel-by-pixel based $\beta_\mathrm{c}$--$T_\mathrm{c}$
 relation (filled square) overplotted on the relation for individual
 dwarf galaxies from the  Dwarf Galaxy Survey with \textit{Herschel}
 (filled circles). The filled square is obtained from
 Figs.~\ref{fig7}(b) and (d) with the SPIRE 500~$\mu$m-band PSF. 
$\beta_\mathrm{c} \ga 1.7$ is found in the western arm. The
 black-filled circle denotes the plot for the whole region of
 NGC~1569. }  
\label{fig9}
\end{figure*}

Figure~\ref{fig9} shows the $\beta_\mathrm{c}$--$T_\mathrm{d}$ relation obtained from
Figs.~\ref{fig7}b and \ref{fig7}d (filled square)
and from individual dwarf galaxies observed by Herschel~\citep{Remy2013,
Remy2015}. Although \cite{Remy2013} investigated 
$\beta$ and $T_\mathrm{d}$ values for the dwarf galaxy samples, they 
used the data sets processed with an older calibration version than our version.
Thus, we reanalyzed $\beta$ and $T_\mathrm{d}$ values for the samples
using flux densities shown in \cite{Remy2015}, which were calibrated with 
the latest version, by applying the same SED fitting procedure as described in
\cite{Remy2013}. The distribution of our data points overlaps well with
the $\beta$-$T_\mathrm{d}$ anti-correlation from individual dwarf galaxies
observed by Herschel~\citep{Remy2013, Remy2015}. Furthermore, our data
plots are in agreement with that for the whole galaxy of
NGC~1569~(black-filled circle).

\begin{figure*}
\centering
 \includegraphics[width=12cm, clip]{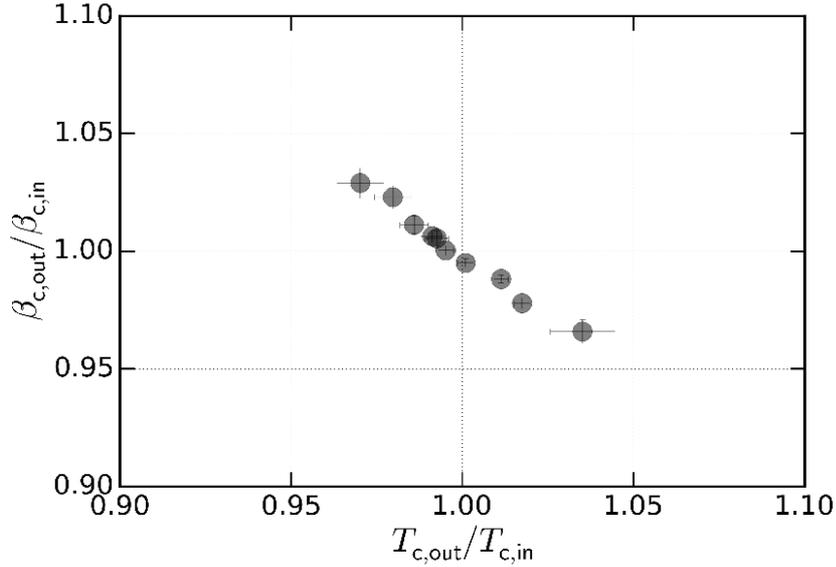}
\vspace{-3mm}
 \caption{Output/input ratio of $\beta_\mathrm{c}$ and $T_\mathrm{c}$
 based on SED fitting of the mock data sets with $\beta_\mathrm{w}=2.0$. Each
 filled circle shows the average of
 ($\beta_\mathrm{c,out}/\beta_\mathrm{c,in}$,
 $T_\mathrm{c,out}/T_\mathrm{c,in}$) sets in each $\beta_\mathrm{c}$
 bin: 10 bins for the $\beta_\mathrm{c}$ range of 1.0--2.0.}   
\label{fig88}
\end{figure*}

To check if the observed $\beta_\mathrm{c}$--$T_\mathrm{d}$ relation
comes from the artificial anti-correlation, we compared input
($\beta_\mathrm{c,in}$, $T_\mathrm{c,in}$) with output
sets of ($\beta_\mathrm{c,out}$, $T_\mathrm{c,out}$) by SED fitting of
the mock data sets. We generated uniform
samples of 2000 random sets of the seven-SED parameters; each
parameter value is uniformly distributed within the range of 5\% of
the original best-fit parameter value to construct realistic mock SEDs.  
Then, each random set of the seven-SED
parameters is used to synthesize flux densities for
$\beta_\mathrm{w}=2.0$ given by Eq.~(\ref{eq1}) at each
wavelength. Finally, the same SED fitting procedure 
as described in Sect.~3.3.1 was applied to the mock
data sets. Figure~\ref{fig88} shows output/input ratio of $\beta_\mathrm{c}$
and $T_\mathrm{c}$ after taking average in each $\beta_\mathrm{c}$ bin
(10 bins for the $\beta_\mathrm{c}$ range of 1.0--2.0). In this figure, although the
artificial anti-correlation is confirmed, the range of both
$\beta_\mathrm{c}$ and $T_\mathrm{c}$ variations is within 
$\sim3$\%. Therefore, the observed $\beta_\mathrm{c}$--$T_\mathrm{d}$
relation is not significantly influenced by the artificial
anti-correlation. Moreover, the enhanced $\beta_\mathrm{c}$ values
reflect variations in the shape of the cold dust component in observed
SEDs.

\begin{figure*}
\centering
 \includegraphics[width=13cm, clip]{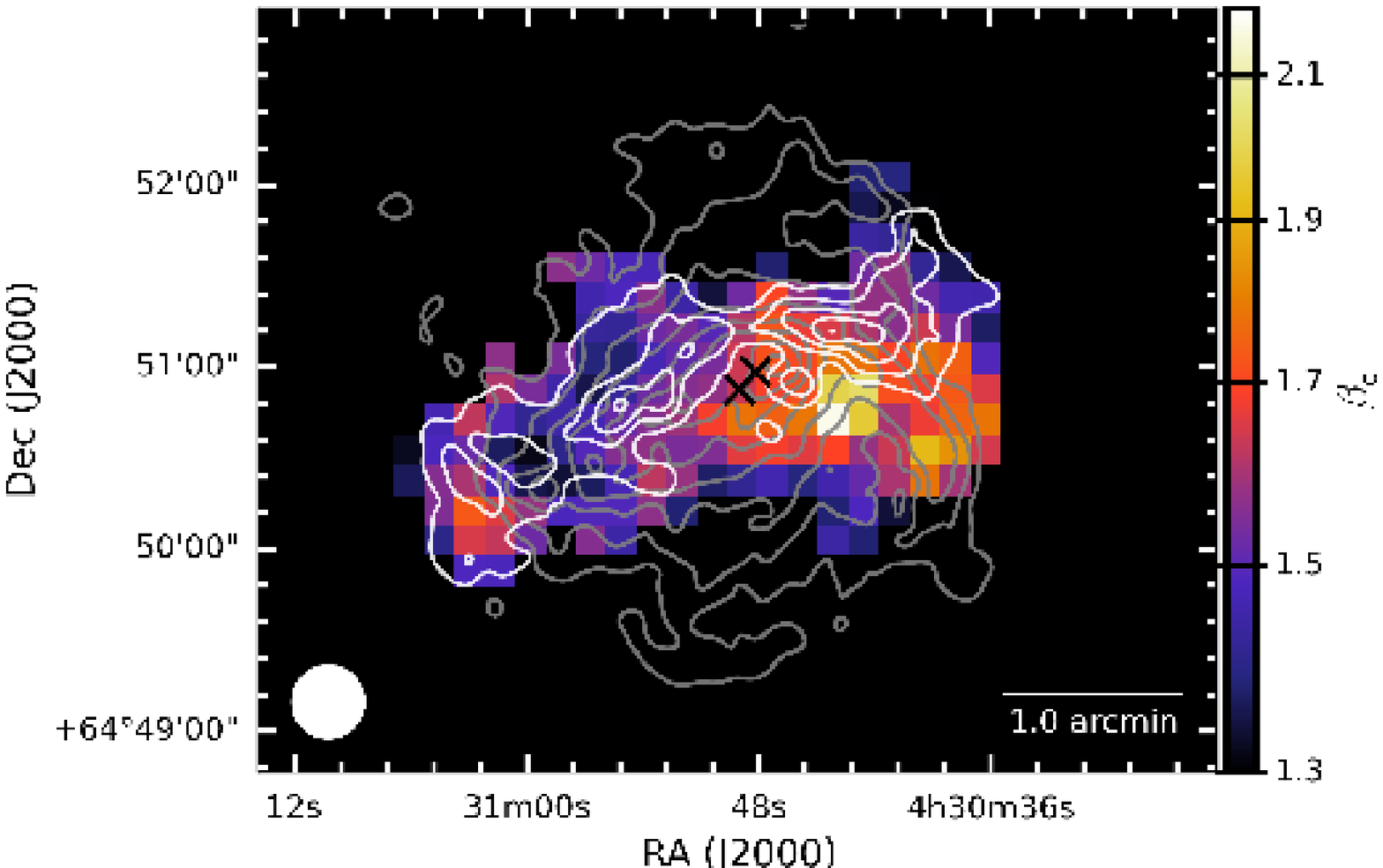}
 \caption{Same as Fig.~\ref{fig77}b but different color scale to show
 regions with $\beta_\mathrm{c}\sim2$. The white contours superposed on  
 the image show the integrated HI~\citep{Walter2008}, while the gray
 contours are the same as those in Fig.~\ref{fig2}.  Two cross marks are
 the positions of SSC-A (right) and B (left).}
\label{fig11}
\end{figure*}

\section{Discussions}
\subsection{Enhanced $\beta_\mathrm{c}$ values along the western arm}
H$\alpha$ observations reveal a number of filaments excited by shocks,
which are associated with the four expanding
superbubbles~\citep{Westmoquette2008}. Moreover, the superbubbles are
filled both with dust grains (see Fig.~\ref{fig2}) and hot plasma~\citep{Martin2002}. 
Among those H$\alpha$ filaments, dust grains associated with the western
arm clearly show enhanced $\beta_\mathrm{c}$ values. Why are such grains
observed along the western arm only?

Theoretically, the Lorentz model gives $\beta$=2.0 in far-IR and sub-mm
region for perfect ionic crystal grains such as crystalline
silicate dust. \cite{Meny2007} proposed a model for dust emission in far-IR
and sub-mm region based on physical properties of amorphous material
(e.g., amorphous silicate). They introduced a disordered charge
distribution combined with the presence of two level tunneling states
to explain a wavelength and temperature-dependent $\beta$. The model
with standard parameters well reproduced Galactic dust emission that
mainly comes from amorphous silicate grains~\citep{Paradis2011}; the
model with the parameters predicts $\beta<2.0$ for the dust
temperature of $\sim30$~K, which is the average $T_\mathrm{c}$
in NGC~1569. 

The fact that $\beta_\mathrm{c}\sim2$ is continuously observed from the 
region around the SSCs to the western arm may attribute to starburst
activities in the SSCs. Figure~\ref{fig11} shows the $\beta_\mathrm{c}$
map together with HI integrated contours~\citep{Walter2008}. In the
disk, the values of $\beta_\mathrm{c}\sim2$ are distributed around the
SSCs (two cross marks), and the region is clearly confined by two dense
HI clouds called as the HI ridge~\citep{Israel1990, Mule2005}. Given
that the SSCs are located in between these two dense HI clouds, as
pointed out by \cite{Johnson2012}, the SSCs are likely to be formed out
of the HI ridge that used to connect these two dense HI
clouds. Those spatial coincidences may indicate that
dust grains around the SSCs are crystalline ones produced by
massive stars originating from starburst activities as have been found
in ultra-luminous infrared galaxies~\citep{Spoon2006}, since the latest
starburst phase ended 5--10~Myr ago~\citep{Angeretti2005}, which is
shorter than the crystalline-to-amorphous conversion timescale of
40--70~Myr due to cosmic-ray hits~\citep{Spoon2006, Kemper2011}. 
This might indicate that crystalline grains injected by massive stars
are blown away along the HI ridge and thus the western arm.

\subsection{Origins of dust heating in the IR-halo region}
In the IR-halo region, spatial distributions of $T_\mathrm{c}$ and
$T_\mathrm{w}$ are asymmetric with respect to the disk.
Two possible heating sources are considered. One is radiative heating
by stars distributed over the disk. In this case, the observed dust
temperatures should monotonically decrease away from the disk.
Therefore, the observed dust temperatures and their distributions in the IR-halo region
cannot be explained solely by radiative dust heating. This means that an
additional heating source is required.  

The other heating source is collisional heating mainly with electrons
in a hot plasma; the possibility is also pointed out
for dust heating in the halo region of NGC~253
by~\cite{kaneda2009}. \cite{Dwek1987} presented a detailed analysis 
of collisional heating of dust grains under various plasma conditions
with temperatures above $\sim 10^6$~K. According to \cite{Dwek1987},
dust temperatures due to collisional heating were calculated by taking
the following conditions into account; for the calculation of the
collisional heating rate of dust grains, a collision partner is mainly
electrons. As for the cooling rate of 
dust grains, the dust absorption coefficient is applied for silicate and 
carbonaceous grains with their sizes of
0.01--1.0~$\mu$m. Since the average number density and temperature of
electrons in the hot plasma are 0.035~cm$^{-3}$ and $3.51\times10^{6}$~K
\citep{Ott2005}, respectively, the expected dust temperature ranges from 15 to
30~K due to collisional heating. Thus, it is likely that the collisional
heating process significantly contributes to dust heating especially in
the northern IR-halo region.

\section{Conclusions}
 \textit{AKARI} and \textit{Herschel} images at wavelengths from
 7~$\mu$m to 500~$\mu$m show a diffuse IR emission extending
 from the galactic disk into the halo region. The most prominent
 filamentary structure seen in the diffuse IR emission is spatially in
 good agreement with the western arm seen in the
 H$\alpha$. The spatial distribution of the 
$F_\mathrm{350}/F_\mathrm{500}$ map shows high values in regions around
 the SSCs and towards the western arm, which are not found in the
 $F_\mathrm{250}/F_\mathrm{350}$ map. The color-color diagram of
 $F_\mathrm{250}/F_\mathrm{350}$--$F_\mathrm{350}/F_\mathrm{500}$
 indicates enhanced $\beta_\mathrm{c}$ in those regions. From a spectral
 decomposition analysis on the pixel-by-pixel basis, we obtained
 $\beta_\mathrm{c}$, $T_\mathrm{c}$, and $T_\mathrm{w}$ maps; the
 $\beta_\mathrm{c}$ map shows values ranging from $\sim1$ to $\sim2$
 over the whole galaxy. In particular, high
 $\beta_\mathrm{c}$ values of $\sim2$ are observed in the regions
 indicated by the color-color diagram.  
 As for the $T_\mathrm{c}$ and $T_\mathrm{w}$ maps, those show high
 temperatures on the northern side than on the southern side in the
 IR-halo region.   

 Since the average cold dust temperature in NGC~1569 is $\sim30$~K,
 $\beta_\mathrm{c}<2.0$ in the far-IR and sub-mm region theoretically
 suggests thermal emission from amorphous grains, while $\beta_\mathrm{c}=2.0$
 suggests that from crystal grains. Given that the enhanced $\beta_\mathrm{c}$
 regions are spatially confined by the HI ridge that is considered to be
 a birthplace of the SSCs, the spatial coincidences may indicate that
 dust grains around the SSCs are crystalline ones injected by massive
 stars originating from starburst activities as have been found in
 ultra-luminous infrared galaxies and that those grains are
 blown away along the HI ridge and thus the western arm.

 The observed asymmetric temperature distribution with respect to the
 disk cannot be explained solely by radiative dust heating by stars
 distributed over the disk. Given that the presence of the hot plasma
 in the IR-halo region, it is likely that the collisional heating
 process significantly contributes to dust heating especially in the
 northern IR-halo region.

\section*{Acknowledgments}
This research is based on observations with AKARI, a JAXA project with
the participation of ESA. 

PACS has been developed by a consortium of institutes led by MPE
(Germany) and including UVIE (Austria); KU Leuven, CSL, IMEC (Belgium);
CEA, LAM (France); MPIA (Germany); INAF-IFSI/OAA/OAP/OAT, LENS, SISSA
(Italy); IAC (Spain). This development has been supported by the funding
agencies BMVIT (Austria), ESA-PRODEX (Belgium), CEA/CNES (France), DLR
(Germany), ASI/INAF (Italy), and CICYT/MCYT (Spain). 

SPIRE has been developed by a consortium of institutes led by Cardiff
University (UK) and including Univ. Lethbridge (Canada); NAOC (China);
CEA, LAM (France); IFSI, Univ. Padua (Italy); IAC (Spain); Stockholm
Observatory (Sweden); Imperial College London, RAL, UCL-MSSL, UKATC,
Univ. Sussex (UK); and Caltech, JPL, NHSC, Univ. Colorado (USA). This
development has been supported by national funding agencies: CSA
(Canada); NAOC (China); CEA, CNES, CNRS (France); ASI (Italy); MCINN
(Spain); SNSB (Sweden); STFC, UKSA (UK); and NASA (USA). 

This work made use of THINGS, 'The HI nearby Galaxy Survey'~\citep{Walter2008}.


%\bibliographystyle{mnras}

\bibliographystyle{mnras}
\bibliography{my_ref}

%%%%%%%%%%%%%%%%%%%%%%%%%%%%%%%%%%%%%%%%%%%%%%%%%%

% Don't change these lines
\bsp	% typesetting comment
\label{lastpage}
\end{document}